\documentclass[pre,showpacs,twocolumn,superscriptaddress]{revtex4}
\usepackage{amssymb,graphicx,amsbsy}

\begin{document}
\title{Critical condition of the water-retention model}
\author{Seung Ki Baek}
\affiliation{Integrated Science Laboratory, Ume{\aa} University, 901 87
Ume{\aa}, Sweden}
\altaffiliation[Present address: ]{School of Physics, Korea Institute for
Advanced Study, Seoul 130-722, Korea}
\author{Beom Jun Kim}
\email[Corresponding author, E-mail: ]{beomjun@skku.edu}
\affiliation{BK21 Physics Research Division and Department of Physics,
Sungkyunkwan University, Suwon 440-746, Korea}

\begin{abstract}
We study how much water can be retained without leaking through boundaries
when each unit square of a two-dimensional lattice is randomly assigned a block
of unit bottom area but with different heights from zero to $n-1$.
As more blocks are put into the system,
there exists a phase transition beyond which
the system retains a macroscopic volume of water.
We locate the critical points and verify that the criticality
belongs to the two-dimensional percolation universality class.
If the height distribution can be approximated as continuous for large $n$,
the system is always close to a critical point and the fraction of the area
below the resulting water level is given by the percolation threshold.
This provides a universal upper bound of
areas that can be covered by water in a random landscape.
\end{abstract}

\maketitle

Percolation is a simple yet powerful model used to understand geometric critical
phenomena~\cite{stauffer,grimmett}. Studies of gelation~\cite{flory} and
fluid in a porous media~\cite{broad} go back more than half a century ago.
Since then, percolation has been studied in a variety of
contexts such as epidemiology~\cite{frisch}, ferromagnetism~\cite{fk},
and even microprocessor manufacturing~\cite{greene}. Recently, it has been
pointed out that percolation also may be studied in a geographic context,
that is, in determining
how much water can be retained without leaking through boundaries
on a given rugged landscape~\cite{ziff}. If the landscape has 
only two levels, the connection to percolation is straightforward. Suppose
that we have a square lattice of a certain size. We may define this initial
state as level zero. It is more intuitive to start with level one, however,
by filling every square site with a cubic block of the unit height. Then, we
randomly remove a certain fraction of the cubic blocks, say $p_0$, so that
we have small ponds here and there where water is retained. As $p_0$
grows, the ponds merge into large clusters, and when one of these clusters
touches the boundary, a large amount of water drains out of the system all at
once. Obviously, this crisis happens when $p_0$ reaches the site-percolation
threshold of the square lattice, which has been estimated as $p_c^{\rm site}
\approx 0.592~746~02(4)$~\cite{site}; in terms of $p_1$, the fraction of
blocks with height $1$, this corresponds to $p_1 = 1 - p_c^{\rm site} =
0.407~253~98(4)$, which we simply denote as $p_c$ throughout this work.
Beyond this direct connection, the water-retention model allows the number
of levels to be larger than $2$, so we may ask ourselves how these different
heights can affect the critical behavior of the system.
The distribution of different heights
extends the parameter space of the original percolation problem to higher
dimensions and thus literally adds a new dimension to percolation, even
though these new parameters turn out to be irrelevant in the
renormalization-group sense (see below). It is
also notable that the analogy of a random landscape can be found in surface
roughening~\cite{fractal} as well as in the concept called a rugged energy
landscape used in spin glasses and protein dynamics~\cite{rugged,protein}.

In the present work we consider this water-retention problem for a general
$n$-level case where the system may have blocks with heights from $0$ to
$n-1$. The fraction $p_i$ of blocks of the height $i$ should satisfy
$\sum_{i=0}^{n-1} p_i = 1$ ($p_i \ge 0$), which makes us deal with  an $(n-1)$-dimensional
parameter space.
A site occupied by a zero-height block is an empty site.
We locate critical points in this parameter space and give numerical
evidence that the criticality always belongs to the two-dimensional
(2D) percolation universality class described by critical
exponents $\beta = 5/36$ and $\nu = 4/3$~\cite{nijs}. The critical condition
implies that one can find a universal feature in any random landscape with
very large $n$, which is largely independent of the distribution of blocks
$\{ p_i \}$. 

Our simulation code is based on the burning algorithm. Let us begin with the
two-level case for ease of explanation. We have an $(L+2) \times (L+2)$ square
lattice, where we assign the boundary state to the outermost squares
while the level-zero state is assigned to the other interior sites. At
each of these $L \times L$ interior sites, we replace its state with a
level-one state with probability $p_1$. This step determines the configuration
of blocks that we are going to examine in the following way.
We pick up a level-zero site from which we run the burning algorithm. 
The fire can neither penetrate
level-one sites nor propagate back. If this fire touches any 
boundary-state site, the water will drain out.
If the burning stops before touching the boundary by being
surrounded by level-one blocks, this burned area is added to the amount of
water retained by this block configuration.
Then we move to another
level-zero site that is not burned yet, which we call available, and repeat
the same procedure until there are no more such available sites left.
One can choose either the depth-first search and the breadth-first search
for the burning algorithm. Although the former exhausts memory faster, it is
simpler to implement  and turns out to have little difficulty in simulating $L
\sim 6 \times 10^3$ using Intel Core2 Duo CPU E8400 with 3.00 GHz and 2.0 GB
memory. The breadth-first search is reserved for very-large-$n$
cases (see below).
Solving a system with blocks of height $0, 1$, and $2$ requires the system to
be burned twice: one occurs at level $1$ and the other at level $2$.
The first burning fills level-zero sites with water up to height $1$
precisely as in the previous two-level case.
After the first burning, consequently, we are left with sites
with levels one and two.
The level-two sites must be occupied by blocks of height $2$,
while the level-one sites can be either occupied by blocks of height $1$ or
filled with water of that height. When we start the second burning,
all these level-one sites should appear available so that we can fill these
sites with water of that height. In other words, only blocks with height $2$
are of our interest at this second burning.
It is straightforward to extend the same procedure to higher levels by
raising the water level one by one.
Note that we count only such water that does not touch the boundary as
retained by the given block configuration every time we run the burning
algorithm.

\begin{figure}
\includegraphics[width=0.45\textwidth]{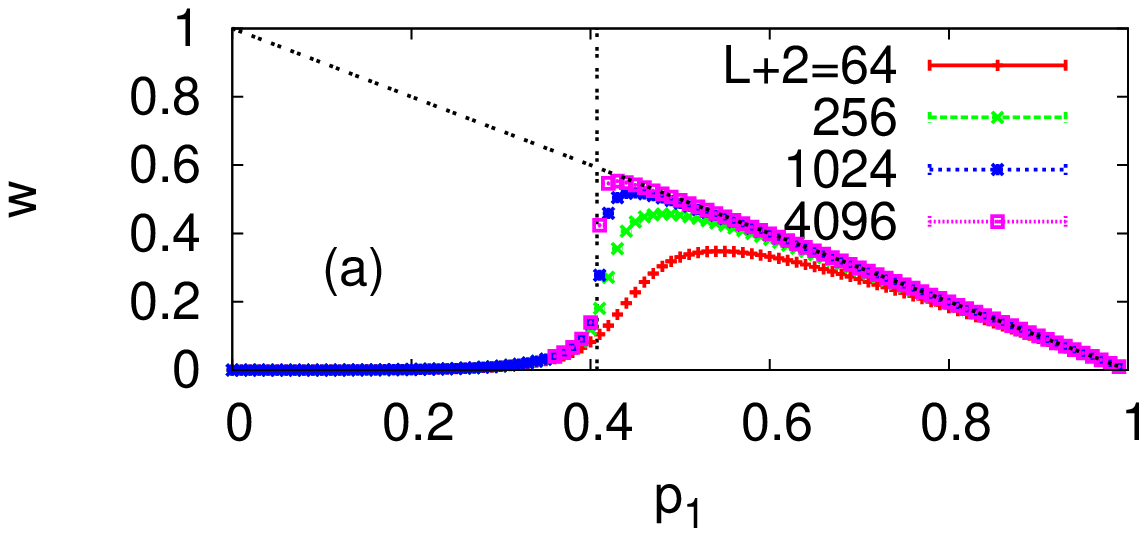}
\includegraphics[width=0.45\textwidth]{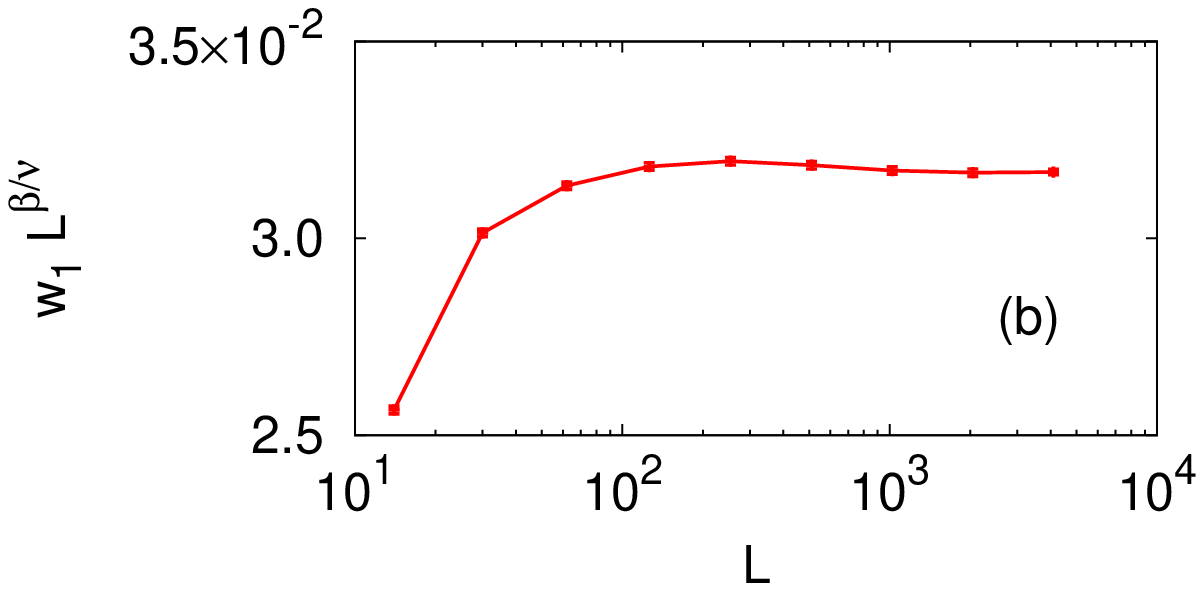}
\caption{(Color online) Results of the two-level case. (a) Water volume per
area $w$ as a function of $p_1$, the fraction of blocks with height $1$.
The vertical dotted line means $p_c \approx 0.407~253~98$
and the other line from the top left to the bottom right is for comparing
$w$ with $1-p_1$. (b) $w_1 L^{\beta/\nu}$ at $p_c$ as a function of $L$,
based on Eq.~(\ref{eq:fss}), where the number of sample averages is
$O(10^5)$ for each data point.
Error bars are shown, but they are comparable to the symbol size.
}
\label{fig:single}
\end{figure}

We start from the two-level case of the water-retention problem~\cite{ziff}.
As discussed above,
it is easier to understand this problem if we start from
$p_1 = 1$ and decrease it by adding level-zero sites.
One may expect that the amount of water will be roughly proportional
to $p_0$, especially for $p_0 \ll
p_c^{\rm site}$, as shown in Fig.~\ref{fig:single}(a). As the
system size grows, the linear proportionality becomes more and more
accurate because the majority of ponds are centrally located away from
the boundary. In Fig.~\ref{fig:single}(a)
the vertical axis means water volume divided by the total interior area
$L \times L$ and we denote this observable as $w$.
When $p_0$ reaches $p_c^{\rm site}$ or, equivalently, when $p_1$ reaches
$p_c$, percolation occurs and a large amount of water drains out through the
boundary.
In terms of percolation theory, the observable $w$ corresponds to the sum of
sizes of all nonpercolating clusters of level-zero sites in bulk, which
will converge to $p_c^{\rm site}$ in the thermodynamic
limit~\cite{stauffer}.
According to the theory of finite-size scaling~\cite{stauffer},
the largest amount of water in a single pond,
denoted by $w_1$, is expected to be described
by the following form:
\begin{equation}
w_1 = L^{-\beta/\nu} f[(p_1-p_c)L^{1/\nu}],
\label{eq:fss}
\end{equation}
with $\beta = 5/36$ and $\nu = 4/3$ near criticality~\cite{nijs}.
This is readily confirmed by Fig.~\ref{fig:single}(b), where we plot $w_1
L^{\beta/\nu} \rightarrow {\rm const}$ at $p_1 = p_c$.

\begin{figure}
\includegraphics[width=0.21\textwidth]{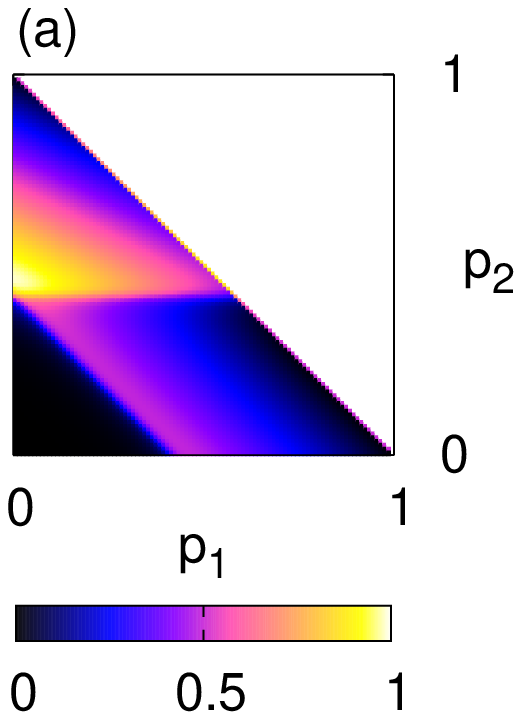}
\includegraphics[width=0.21\textwidth]{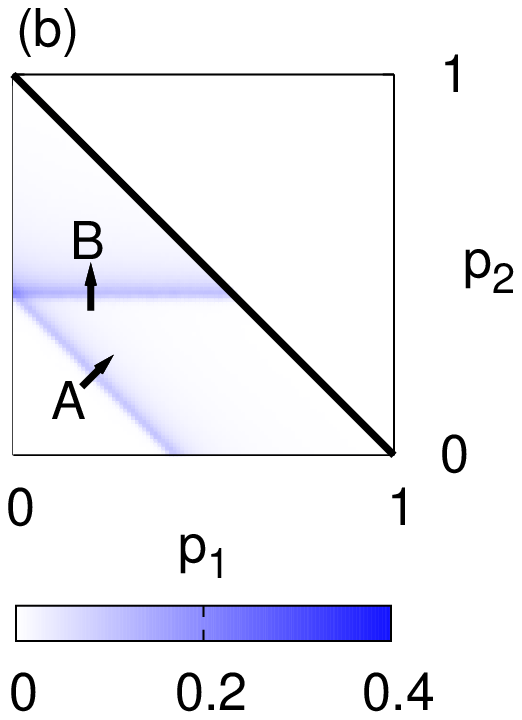}
\includegraphics[width=0.45\textwidth]{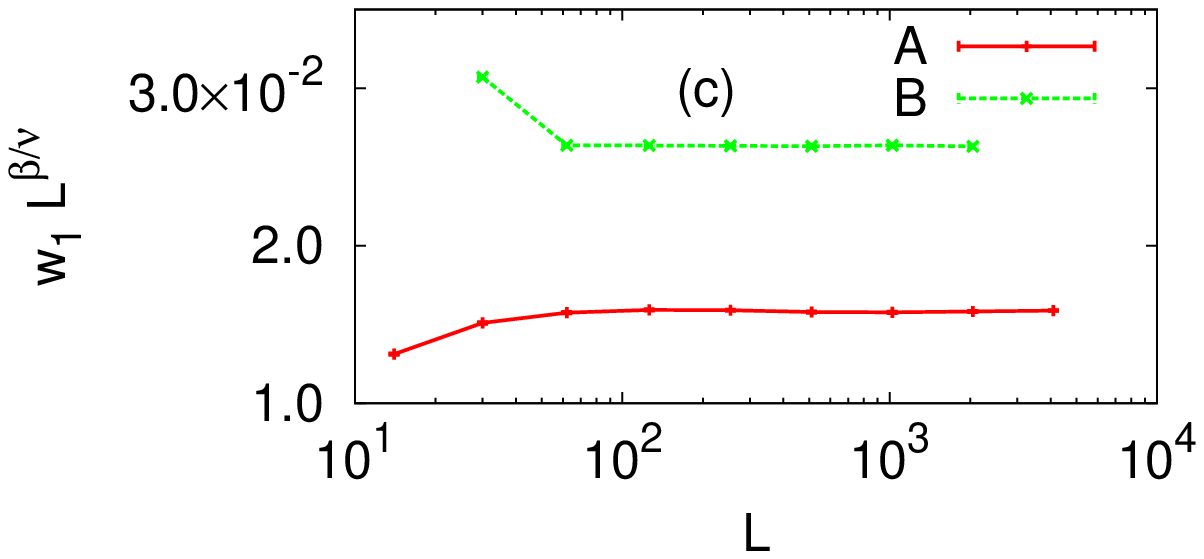}
\caption{
(Color online) Results of the three-level case. (a) $w$ as a
function of $p_1$ and $p_2$ at $L+2=512$. (b) Difference in $w$ between
$L+2=256$ and $512$. Cross section $A$, depicted as a slanting arrow,
means $p_1=p_2$, and cross section $B$, depicted as a vertical arrow,
means $p_1=p_c/2$. (c) $w_1 L^{\beta/\nu}$ measured at
$p_1=p_2=p_c/2$, the critical point along $A$, and also at $p_1 = p_2/2 =
p_c/2$, the critical point along $B$. The 2D percolation result
$\beta/\nu=5/48$ is used in common.
}
\label{fig:three}
\end{figure}
In the three-level case the result can be depicted in a triangular
area restricted by $p_1+p_2 \le 1$, as shown in Fig.~\ref{fig:three}(a), since
the fraction of level-zero sites is obviously $p_0=1-p_1-p_2$ and $p_0 \ge 0$. 
At each of the three boundaries of this triangle, we are back to the
two-level case. First, if
$p_2=0$, we move along the low horizontal boundary and the result is
exactly Fig.~\ref{fig:single}(a).
Second, if $p_2=1-p_1$, we move along the hypotenuse of the triangle and it
is again the two-level problem with level-one and level-two sites because
$p_0 = 0$.
Finally, if $p_1=0$, we move along the left vertical boundary. It is the
same pattern as in the two-level case but with double amounts of water since
the block
height is doubled. The question is then what happens inside these
boundaries; Fig.~\ref{fig:three}(a) gives the answer.

Let us locate critical points inside the triangle. For this
purpose we compare two different system sizes in Fig.~\ref{fig:three}(b).
The motivation is as follows. When there are only small ponds around us,
one hardly needs to care about how large the system is because everything is
local. If the system is close to the critical point so that a large amount of
water is about to drain out, however, the correlation length becomes
so large that the behavior is essentially constrained by the finite size of the system.
In other words, it is around the critical point that our observable $w$
becomes the most sensitive to the system size, as can be seen in
Fig.~\ref{fig:single}(a) for the two-level case.
Figure~\ref{fig:three}(b) indicates that there are two lines of critical
points, which can be written as
\begin{eqnarray}
&&p_1 + p_2 = p_c \label{eq:line1}\\
&&p_2 = p_c. \label{eq:line2}
\end{eqnarray}
We can argue that these are exact expressions for the following reason:
Eq.~(\ref{eq:line1}) implies that blocks with height $2$ can be substituted
for those
with height $1$, while Eq.~(\ref{eq:line2}) implies that the blocks with
height $2$ have their own critical point without being affected by the
presence of level-one blocks.

Another important question is whether the critical property remains the same
along these critical lines. The above argument on exactness of
Eqs.~(\ref{eq:line1}) and (\ref{eq:line2}) implies that this will be the
case. We examine two cross sections represented as $A$ and $B$ in
Fig.~\ref{fig:three}(b) and
show the results in Fig.~\ref{fig:three}(c). In both these cases
$w_1$ follows the scaling form of Eq.~(\ref{eq:fss}) with $\beta/\nu = 5/48$.
We have also observed that the critical region of $w$ scales as $L^{-1/\nu}$
with $\nu = 4/3$ (not shown here). These results strongly suggest that the
criticality always belongs to the 2D percolation universality class.

\begin{figure}
\includegraphics[width=0.21\textwidth]{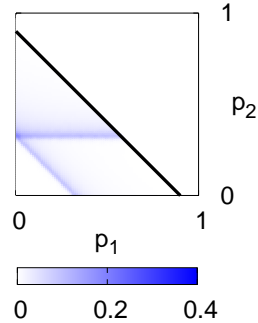}
\caption{(Color online) Results of the four-level case. The calculation is
the same as Fig.~\ref{fig:three}(b) except that $p_3 = 0.1$.}
\label{fig:four}
\end{figure}

A four-level system requires consideration of the volume inside a
tetrahedron, written as $p_1 + p_2 + p_3 \le 1$. Since it is not easy to
visualize such a three-dimensional object, we plot differences in $w$
between two system sizes as before, but now fixing $p_3$ at $10\%$ in
Fig.~\ref{fig:four}. Comparing this with Fig.~\ref{fig:three}(b), we
immediately see that the new critical surfaces can be written as
\begin{equation}
\begin{array}{l}
p_1 + p_2 + p_3 = p_c,\\
p_2 + p_3 = p_c,\\
p_3 = p_c.
\end{array}
\label{eq:four}
\end{equation}
Note that we had $p_1 = p_c$ as the critical condition for the two-level case
and Eqs.~(\ref{eq:line1}) and (\ref{eq:line2}) for the three-level case.
Comparing Eq.~(\ref{eq:four}) with these previous cases, we can now
generalize the critical condition for the $n$-level case as
\begin{equation}
\begin{array}{l}
P(1) = p_c,\\
P(2) = p_c,\\
~~~\vdots\\
P(n-1) = p_c,
\end{array}
\label{eq:gen}
\end{equation}
where $P(i) \equiv \sum_{j=i}^{n-1} p_j$. Note that the system is
critical when {\em any} of the conditions is satisfied. Let us assume that
$P(i)$ can be approximated by a continuous function of $i$ as $n$ gets large.
It implies that $p_i$ can be made arbitrarily small by dividing levels ($n
\rightarrow \infty$),
which is true for most natural landscapes where contour lines do not occupy a
finite area fraction. Then one can argue from the
intermediate-value theorem that there exists a certain $i=i^{\ast}$ that
satisfies one of the critical conditions above.
In other words, the system is almost always close to a certain
critical surface so that the correlation length is found to be very large at a
certain water level $i^{\ast}$. Consider a system where a large amount of
water is poured onto the model and the excess water is allowed to drain off.
For $i \ge i^{\ast}$,
water will flow out of the system, so the water level should be
kept slightly below $i^{\ast}$ in a stable situation. Still, $P(i^{\ast})$ is
close to $p_c$ and the typical length scale will be very large, which shapes
observed ponds and lakes into fractal objects like 2D percolating
clusters. In this sense, one may connect this finding to the idea of
self-organized criticality~\cite{soc}, although
our finding originates from gradual variations in $P(i)$,
not from any dynamic process (see also Ref.~\cite{inv} on self-organizied
behavior in invasion percolation).

Another implication of Eq.~(\ref{eq:gen}) is that
the fraction of the area above the observed water level $i^{\ast}$ should be
always the same for infinitely many $(p_1, p_2, p_3, ..., p_{n-1})$.
The fraction of the area below the water level is given by $p_c^{\rm site} =
0.592~746~02(4)$ in our case, but the numeric value depends on the square
geometry we have chosen, so actual field observations may well give a
different value. The important point is that there can be a universal upper
limit of how much area can be covered by water at any length scale. Since we
have considered randomly distributed blocks, if a landscape contains more
water than this upper limit, it indicates the existence of non-random
processes with correlation lengths comparable to the scale of our
observation.

\begin{figure}
\includegraphics[width=0.45\textwidth]{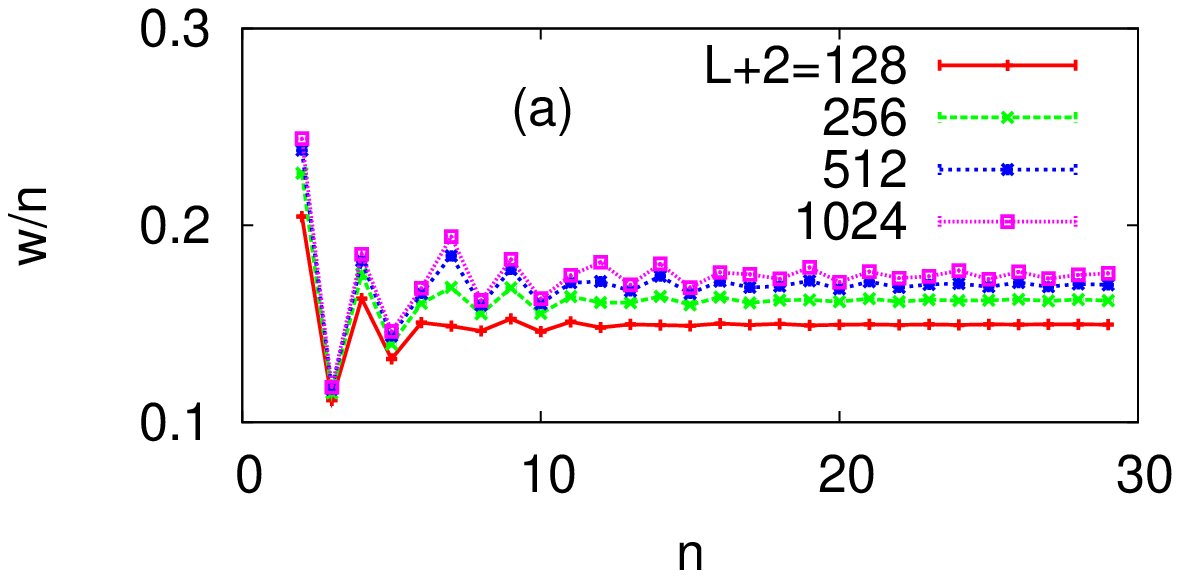}
\includegraphics[width=0.45\textwidth]{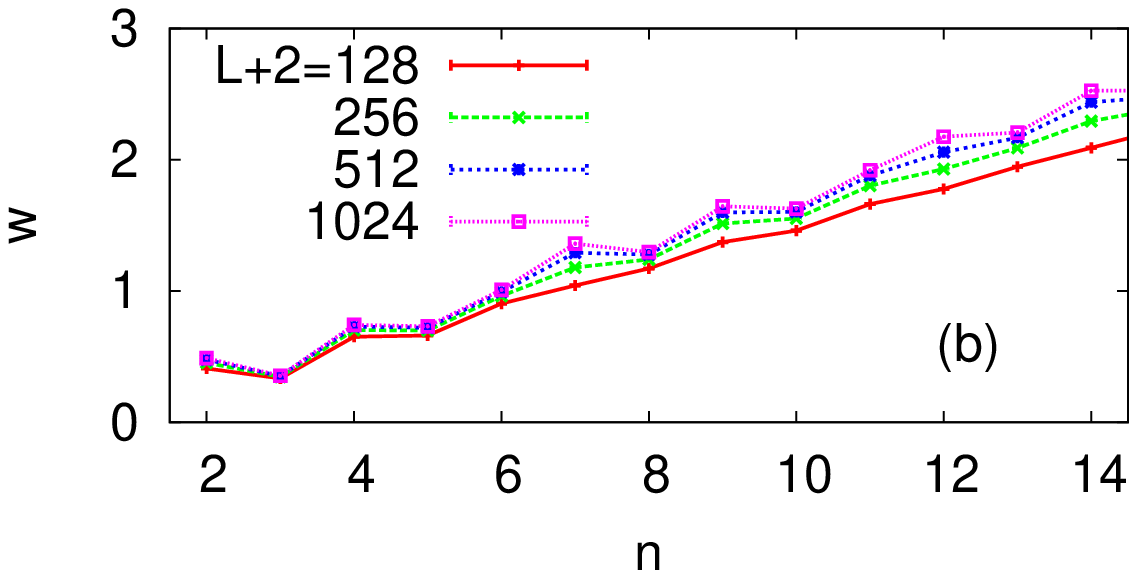}
\caption{(Color online) Retained water as a function of $n$, the number of
levels, where $p_i = 1/n$ for every $i$.
(a) As $L \rightarrow \infty$, $w/n$, with $n \gg 1$, approaches a constant.
(b) For finite $n$, $w$ is not monotonic: $n=3$
contains a smaller amount than $n=2$, for example, and such nonmonotonic
behavior becomes more visible with larger $L$.
}
\label{fig:cross}
\end{figure}

As a simple example, let us assume that $p_i$'s are equally distributed,
i.e., $p_i = 1/n$ for every $i$~\cite{ziff}. When $n$ is very large, we have
$P(i) \approx 1 - i/n$ and thus $P(i^{\ast}) = 1-i^{\ast}/n = p_c$. Water
retained in this system roughly amounts to $w = \sum_{i=0}^{i^{\ast}}
(i^{\ast} - i)p_i \approx (1-p_c)^2 n/2 \approx 0.175~673~9 n$. In fact, this
is an underestimate since there can be ponds with water higher than
$i^{\ast}$~\cite{priv}. The point is that $w$ is asymptotically
proportional to $n$ [Fig.~\ref{fig:cross}(a)]. However, $w$
might not be a monotonic function of $n$ when $n$ is finite
[Fig.~\ref{fig:cross}(b)]. The reason can be found by recalling
Eq.~(\ref{eq:gen}): The question is how close one can get to a critical
condition at each
given $n$ while keeping $P(i)$ larger than $p_c$. For example, one can get
$P(1) = 1/2$ for $n=2$, which differs from $p_c$ by $0.09$. For $n=3$, the
minimal difference gets larger because $P(2) - p_c \approx 2/3 - 0.4 \approx
0.26$. It explains the increasing value of $w(n=2) - w(n=3)$ with increasing
$L$ and the same explanation holds
true for larger $n$'s as well in Fig.~\ref{fig:cross}(b).
 
In summary, we have considered the general $n$-level water-retention problem.
Specifically, we have found the critical conditions
in the $(n-1)$-dimensional parameter space and provided numerical evidence
that the critical property always belongs to the 2D percolation universality
class. The critical condition implies that there are universal features in
the large-$n$ limit: First, the system is very likely to be close to a
critical point and second, the area below water is not
significantly dependent on the distribution of block heights.
This tells us how the percolation threshold
can be related to the upper limit of water retained by a random landscape
and used as a quantitative measure for detecting a geophysical process
on a certain length scale when its existence is called into question.\footnote{
It is interesting to recall that $70.9\%$ of the Earth is covered by
water~\cite{cia}. This fraction is higher than site-percolation thresholds
of the triangular~\cite{sykes}, square~\cite{site}, and honeycomb
lattices~\cite{gu} and also higher than critical area fractions of 2D
continuum percolation models~\cite{cont1,cont2}. If we assume that every
geophysical process has a short length scale compared to the size of the
Earth, so that the global landscape on the Earth can be roughly regarded as
random, we may retrospectively understand that too much of the Earth is covered
by water to be believed to be a flat disk with steep cliffs at the boundary
as once believed.}

\acknowledgments
We are deeply indebted to Craig L. Knecht for introducing us to this problem,
and communication with Robert M. Ziff is gratefully acknowledged.
S.K.B. acknowledges the support from the Swedish Research Council with
Grant No. 621-2008-4449. B.J.K. was supported by Faculty Research Fund,
Sungkyunkwan University. This research was conducted using the
resources of High Performance Computing Center North.


\end{document}